\documentclass[twocolumn,secnumarabic,
reprint,
nobibnotes,
 amsmath,amssymb,
 aps,prb
]{revtex4-2}

\usepackage{graphicx}
\usepackage{dcolumn}
\usepackage{bm}
\usepackage[dvipsnames,table,xcdraw]{xcolor}
\usepackage{booktabs}

\usepackage{comment}

\begin{document}

\preprint{APS/123-QED}

\title{
Axionic $p+is$ superconductivity from two pairing channels
}

\author{Maitê Kessler de Azambuja
} 
\author{David  M\"{o}ckli}
\affiliation{Instituto de F\'{i}sica, Universidade Federal do Rio Grande do Sul, 91501-970 Porto Alegre, Brazil}

\date{\today}

\begin{abstract}
In unconventional superconductors, inversion and/or time-reversal symmetries may be broken, either extrinsically or spontaneously.
Here, we consider a simple BCS model with both attractive singlet and attractive triplet pairing channels. 
We show that when the triplet instability dominates, the model predicts an axionic superconducting state, in which both inversion and time-reversal symmetries are spontaneously broken by the superconductivity without requiring spin-orbit coupling. This leads to characteristic experimental signatures of spontaneous symmetry breaking in superconductors, such as a two-step transition in the specific heat. 
We critically analyze whether familiar pairing mechanisms such as the electron-phonon interaction and ferromagnetic spin fluctuations could produce such an axionic state.
\end{abstract}

\maketitle


\section{Introduction}
Inversion and time-reversal symmetries are both present in BCS theory. Some unconventional superconductors break both symmetries \cite{Davydova2024}.
The inversion symmetry breaking may be inherited from the normal state, such as in noncentrosymmetric superconductors \cite{Sigrist2009,bauer2012,Yip2014}. 
Yet, the joint symmetry breaking is also possible in inversion-symmetric crystals if both inversion and time-reversal symmetries are spontaneously broken by the superconducting state itself \cite{Watanabe2022}.

In systems where inversion (parity, $\mathcal{P}$) and time-reversal ($\mathcal{T}$) breaking occur spontaneously in the superconducting state, the combined symmetry $\mathcal{PT}$  may still be preserved \cite{Watanabe2022,Kanasugi2022}.
This may occur in multiband systems due to mixed-parity interband pairing, which is referred to as anapole superconductivity \cite{Kanasugi2022}. 
However, a $\mathcal{PT}$-preserving $p+is$ state may also occur in single-band systems with spin-orbit coupling, which is sometimes called \textit{axionic superconductivity} \cite{Goswami2014,Wang2017}.

The topological field theory describing time-reversal-invariant insulators in (3+1) dimensions is given by the axion Lagrangian, originally formulated in the context of quantum chromodynamics \cite{Qi2008}. Refs. \cite{Goswami2014,Wang2017} demonstrated the existence of condensed matter analogs of axion electrodynamics, among which the parity-mixed $p+is$ superconductor serves as a prominent example.

Here, we consider a simple BCS-type theory in which both inversion and time-reversal symmetries are present in the normal state. Unlike the original BCS theory, which contains only an attractive singlet channel \cite{Bardeen1957}, we also include an attractive triplet channel. We show that when the triplet channel dominates, this results in a spontaneously broken axionic $p + i s$ state. This demonstrates axionic superconductivity without requiring spin-orbit coupling or multiband effects, while still exhibiting experimental signatures characteristic of  chiral superconductors. As an illustration, we analyze the behavior of the specific heat.

One might then think that triplet-dominated two-channel superconductivity provides the minimal ingredients for an axionic superconductor. However, triplet superconductors are hard to come by, and although theoretical approaches frequently assume the existence of two attractive channels \cite{Sigrist1991,Goswami2014,mockli2019}, demonstrating such systems from the perspectives of fermiology, energetics, and pairing mechanisms is considerably more challenging. For this reason, we examine the Hubbard-Holstein model as a candidate to support the two-channel BCS model in the weak-coupling limit. Our analysis shows that, under standard conservative assumptions, this model does not generically produce axionic superconductivity, thus complicating the search for realistic axionic superconducting systems. 

We organize the paper as follows. In Sec. \ref{sec:two_channel}, we formulate the two-channel BCS model containing both singlet and triplet instabilities. In Sec. \ref{sec:gl}, we derive the corresponding Ginzburg–Landau free energy, which allows us to study the temperature dependence of the superconducting order parameters, the symmetries, the phase diagram, and the specific heat. In Sec. \ref{sec:mechanisms}, we discuss whether the interactions present in the Hubbard–Holstein model produce axionic superconductivity. We conclude in Sec. \ref{sec:discussion}.

\section{Two channel BCS model \label{sec:two_channel}}

Consider the generalized BCS Hamiltonian of the form \cite{Sigrist1991,Sigrist2005}
\begin{align}
    & \mathcal{H} = \sum_{\boldsymbol{k},s}\xi_{\boldsymbol{k}}c_{\boldsymbol{k}s}^\dag c_{\boldsymbol{k}s} \notag \\
    &+
    \frac{1}{2}\sum_{\boldsymbol{k},\boldsymbol{k}',\{s_i\}}V_{s_1s_2,s_1's_2'}(\boldsymbol{k},\boldsymbol{k'})
c^\dag_{-\boldsymbol{k}s_1}c^ \dag_{-\boldsymbol{k}s_2}c_{-\boldsymbol{k}'s_2'}c_{\boldsymbol{k}'s_1'},
\label{eq:bcs}
\end{align}
where $\xi_{\boldsymbol{k}} = \xi_{-\boldsymbol{k}}$ denotes the inversion-symmetric normal-state dispersion, and $c_{\boldsymbol{k}s}^\dag$ ($c_{\boldsymbol{k}s}$) are fermionic creation (annihilation) operators with momentum $\boldsymbol{k}$ and spin projection $s=\ \uparrow,\downarrow$.
The superconducting pairing channels are given by 
\begin{align}
& V_{s_1s_2,s_1's_2'}(\boldsymbol{k},\boldsymbol{k'}) = \notag \\
&
+\sum_{\Gamma,j}
v_{s,\Gamma,j}
\left(\hat{\psi}_{\boldsymbol{k},\Gamma_j}i\sigma_y\right)_{s_1s_2}
\left(\hat{\psi}_{\boldsymbol{k}',\Gamma_j}i\sigma_y\right)^ *_{s_1's_2'}  \notag \\
&
+ 
\sum_{\Gamma,j}
v_{t,\Gamma,j}
\left(\hat{\boldsymbol{d}}_{\boldsymbol{k},\Gamma_j}\cdot\boldsymbol{\sigma}i\sigma_y\right)_{s_1s_2}
\left(\hat{\boldsymbol{d}}_{\boldsymbol{k}',\Gamma_j}\cdot\boldsymbol{\sigma}i\sigma_y\right)^ *_{s_1's_2'},
\label{eq:channels}
\end{align}
where $j$ indexes the basis functions ($\hat{\psi}_{\boldsymbol{k},\Gamma_j}$ and $\hat{\boldsymbol{d}}_{\boldsymbol{k}',\Gamma_j}$) of an irreducible representation $\Gamma$, with $v_{s,\Gamma}$ representing the pairing interactions in the singlet channels and $v_{t,\Gamma}$ in the triplet channels. 
Also, $\boldsymbol{\sigma}=(\sigma_x,\sigma_y,\sigma_z)$ is the vector of Pauli matrices in spin space. 
Whereas Eq. \eqref{eq:channels} is a generic decomposition into singlet and triplet channels, 
a specific pairing mechanism modeling populates the channels $\{v_{s,\Gamma,j},v_{t,\Gamma,j}\}$, which might be attractive or repulsive.

BCS theory assumes a single attractive singlet $s$-wave channel, with the critical temperature determined by 
\begin{align}
T_{\mathrm{c}}=\frac{2e^\gamma}{\pi}\omega_\mathrm{C}\,e^{-1/\lambda_s},
\end{align}
where $\gamma$ is Euler's constant, $\lambda_s=N_0v_s$ is the coupling strength, $N_0$ is the density of states at the Fermi level, and $\omega_\mathrm{C}$ is a characteristic cutoff energy that frequently relates to phonons, but may be of generic origin.

We now entertain the possibility of two attractive channels with opposite parities. For discussion purposes, we assume a cubic crystal described by the point group $O_h$. This allows us to maintain the discussion simple, without  
relying on phenomena that would require specialized point groups. 
Within $O_h$, the attractive even-parity singlet $s$-wave channel is $v_{s,A_1}=v_s<0$ and $\hat{\psi}_{\boldsymbol{k},A_1}=1$. 
For $O_h$, the 3D irrep of the odd-parity triplet channel is $T_{1u}$, where we use $v_{t,T_{1u},j}=v_t<0$, with basis functions $\hat{\boldsymbol{d}}_{\boldsymbol{k},T_{1u,j}}$ corresponding to $p$-waves in the three spatial directions.

In contrast to single channel BCS theory, we now have
\begin{align}
& T_{\mathrm{cs}}=\frac{2e^\gamma}{\pi}\omega_\mathrm{D}\,e^{-1/\lambda_s},\quad T_{\mathrm{ct}}=\frac{2e^\gamma}{\pi}\omega_\mathrm{F}\,e^{-1/\lambda_t}, \label{eq:tcs}
\end{align}
where $\lambda_t=3N_0v_t$.
Here, $T_{\mathrm{cs}}$ or $T_{\mathrm{ct}}$ is the superconducting transition temperature that the system would have in the absence of the other channel. Whether it is $T_{\mathrm{cs}}$, $T_{\mathrm{ct}}$, or some other $T_\mathrm{c}'$ that corresponds to the true transition temperature, depends on solving the model for a specific set of input parameters. The notation of the cutoff energies $\{\omega_\mathrm{D},\omega_\mathrm{F}\}$ is suggestive that the singlet channel may be of phononic origin, whereas the triplet channel has a purely fermionic cutoff. However, following the same rationale as BCS, here we abstain from associating a specific origin to $\{\omega_\mathrm{D},\omega_\mathrm{F}\}$, and treat them as input parameters instead. Therefore, through Eqs. \eqref{eq:tcs}, we use the pair $\{T_{\mathrm{cs}},T_{\mathrm{ct}}\}$ as input parameters to the two channel BCS theory.

It is known that a superconducting state with coexisting singlet and triplet components breaks parity, and finds a natural environment in noncentrossymetric crystals
\cite{Sigrist2005,bauer2012}. 
But since the normal state of the Hamiltonian in Eq. \eqref{eq:bcs} is inversion and time-reversal symmetric, one would therefore not expect coexistence from symmetry arguments alone, unless parity is broken spontaneously by the superconducting state. 
This would correspond to a $\mathcal{P}$-broken state, and depending on the energetics a $\mathcal{T}$-broken chiral state. 
Materials that have inversion and time-reversal symmetric normal state, but the superconducting state breaks both parity and time-reversal spontaneously, yet retain the $\mathcal{PT}$ symmetry,
are called \textit{axionic superconductors} \cite{Goswami2014,Wang2017,Watanabe2022}. 
Such superconductors would be unconventional, but do not  necessarily rely on spin-orbit coupling, magnetic field, or interfaces. The requirement would be two active pairing channels of opposite parity. 

\section{Ginzburg-Landau theory \label{sec:gl}}

Given the two-channel BCS model with input parameters $\{T_{\mathrm{cs}},T_{\mathrm{ct}}\}$,
we can then consider three cases: (i) the singlet instability dominates ($T_\mathrm{cs} > T_\mathrm{ct}$), (ii) the critical temperatures are accidentally degenerate ($T_\mathrm{cs} = T_\mathrm{ct}$), and 
(iii) the triplet instability dominates ($T_\mathrm{ct} > T_\mathrm{cs}$). 
To evaluate all possible solutions, we derive the weak-coupling free energy expansion for the Hamiltonian in Eq. \eqref{eq:bcs}.

\subsection{Derivation}
We now perform a mean-field approximation on the Hamiltonian in Eq. \eqref{eq:bcs}, such that the order parameters are \cite{Sigrist1991}
\begin{align}
 \Delta_{s_1s_2}(\boldsymbol{k})=\sum_{\boldsymbol{k}',s_1',s_2'}V_{s_1s_2,s_1's_2'}(\boldsymbol{k},\boldsymbol{k}')\left\langle c_{-\boldsymbol{k}'s_2'}c_{\boldsymbol{k}'s_1'} \right\rangle. 
 \label{eq:op}
\end{align}
In spin space, Eq. \eqref{eq:op} may be expressed as
\begin{align}
\Delta(\boldsymbol{k})=\left(\Psi(\boldsymbol{k})\sigma_0+\boldsymbol{d}(\boldsymbol{k})\cdot\boldsymbol{\sigma}\right)i\sigma_y,
\end{align}
where $\Psi(\boldsymbol{k})$ parametrizes the singlet, and the $\boldsymbol{d}$-vector parametrizes the triplet order parameter. These may be expressed in terms of basis functions as
\begin{align}
\Psi(\boldsymbol{k})=
\sum_i^{l_\Gamma}\psi_{\Gamma_i}\hat{\psi}_{\Gamma_i}(\boldsymbol{k}),\quad
\boldsymbol{d}(\boldsymbol{k})=
\sum_{i=1}^ {l_\Gamma}\eta_{\Gamma_i} \hat{\boldsymbol{d}}_{\Gamma_i}(\boldsymbol{k}),
\end{align}
where $l_\Gamma$ is the dimensionality of the irrep, and the $\{\psi_{\Gamma_i},\eta_{\Gamma_i} \}$ serve as the complex valued Ginzburg-Landau order parameters. 
The ($A_{1g}$) $s$-wave order parameter is $\Psi(\boldsymbol{k})=\psi\,\hat{\psi}_{A_{1g}}(\boldsymbol{k})$, where $\hat{\psi}_{A_{1g}}=1$ for an $s$-wave order parameter.
These basis functions are normalized according to 
\begin{align}
\left\langle \left|\hat{\psi}_{\Gamma_i}(\boldsymbol{k})\right|^2 \right\rangle_{\boldsymbol{k}}=1,
\end{align}
where the Fermi surface average $ \left\langle \ldots \right\rangle_{\boldsymbol{k}}\equiv \int\frac{\mathrm{d}\Omega_{\boldsymbol{k}}}{4\pi}(\ldots)$,
and $\Omega_{\boldsymbol{k}}$ is the solid angle.

Since we are working with $T_{1u}$ in $O_h$, we look at the Balian-Werthamer-state (BW) \cite{Balian1963,Sigrist2005}
\begin{align}
 \hat{\boldsymbol{d}}(\boldsymbol{k})=\eta\left(\hat{d}(k_x),\hat{d}(k_y),\hat{d}(k_z)\right),
 \label{eq:bw}
\end{align}
where $\hat{d}(k)$ are normalized odd functions of momentum according to
\begin{align}
 \left\langle \sum_{i=1}^3\left|\hat{d}(k_i)\right|^2 \right\rangle_{\boldsymbol{k}}=1.
 \label{eq:normalization}
\end{align}
The lowest harmonic of $\hat{d}(k)$ is a $p$-wave, such as $\sin(ka)$.

The mean-field theories' corresponding free energy is given by \cite{mockli2019}
\begin{align}
F = & -\frac{1}{2}\sum_{\boldsymbol{k},\boldsymbol{k}'}\sum_{s_i,s_i'}\Delta^*_{s'_1s'_2}(\boldsymbol{k}')V^{-1}_{s_1's_2',s_1 s_2 }(\boldsymbol{k}^\prime,\boldsymbol{k}) \Delta_{s_1 s_2}(\boldsymbol{k}) \notag \\
&
-\frac{1}{\beta}\sum_{\boldsymbol{k},\omega_n} \mathrm{tr}\,\ln\frac{\beta}{2}\left(-\mathcal{G}^{-1}(\boldsymbol{k},\omega_n) \right ),
\end{align}
where the inverse Green's function is
\begin{align}
 \mathcal{G}^{-1}(\boldsymbol{k},\omega_n)=
\begin{bmatrix}
G_0^ {-1}(\boldsymbol{k},\omega_n)\sigma_0 &  \Delta(\boldsymbol{k})\\
\Delta^\dag(\boldsymbol{k}) &  G_0^ {-1,T}(\boldsymbol{k},\omega_n)\\
\end{bmatrix},
\end{align}
with the normal state inverse Green's function $G_0^ {-1}(\boldsymbol{k},\omega_n)=-i\omega_n+\xi_{\boldsymbol{k}}$, $\beta=1/T$ is the inverse temperature, and $\omega_n=(2n+1)\pi/\beta$ are the fermionic Matsubara frequencies, where we implicitly use the Boltzmann constant as $k_\mathrm{B}=1$. 

We expand the free energy up to quartic order in the order parameters to obtain
\begin{align}
&\frac{1}{2N_0}F[\psi,\eta] = 
a_s(T)|\psi|^2+ a_t(T)|\eta|^2 \notag \\
& +\frac{b(T)}{2}
\biggr\langle
\biggr[ 
|\Psi(\boldsymbol{k})|^4 + |\boldsymbol{d}(\boldsymbol{k})|^4+4|\Psi(\boldsymbol{k})|^2|\boldsymbol{d}(\boldsymbol{k})|^2 \notag \\
& +\left(\Psi^2(\boldsymbol{k})\boldsymbol{d}^*(\boldsymbol{k})\cdot \boldsymbol{d}^*(\boldsymbol{k})+\mathrm{h.c.}\right)+ \left|\boldsymbol{d}(\boldsymbol{k})\times\boldsymbol{d}^*(\boldsymbol{k})\right|^2
\biggr\rangle_{\boldsymbol{k}}, 
\end{align}
where the phenomenological coefficients are microscopically provided by 
\begin{align}
    a_s(T)=\ln\left(\frac{T}{T_\mathrm{cs}}\right),\,\, a_t(T)=\ln\left(\frac{T}{T_\mathrm{ct}}\right),\,\,
    b(T)=\frac{7\zeta(3)}{8\pi^2 T^2}. 
\end{align}
It is instructive to analyze the quartic terms within the average. The first and second terms are the quartic terms of the singlet and triplet instability. The third and fourth terms couple the singlet and triplet order parameters, which lacks at the quadratic order because of inversion symmetry in the normal state. The last term includes non-unitary triplet states, which vanishes for Eq. \eqref{eq:bw}.

We now express both $\psi(\boldsymbol{k})$ and $\boldsymbol{d}(\boldsymbol{k})$ in terms of the basis functions, and perform the averages to obtain
\begin{align}
&\frac{1}{2N_0}F[\psi,\eta] = 
a_s(T)|\psi|^2+ a_t(T)|\eta|^2\notag \\
&+\frac{b(T)}{2}
\biggr[C_1|
\psi|^4+C_2|\eta|^4 \notag \\
&\qquad\qquad
+4C_3|\psi|^2|\eta|^2+C_3\left(\psi^2{\eta^*}^2+\mathrm{h.c.}\right)
\biggr],
\label{eq:free_energy}
\end{align}
where $C_1$, $C_2$ and $C_3$ are real constants given by
\begin{align}
C_1= \left\langle \hat{\psi}^4_{A_{1g}}(\boldsymbol{k})\right\rangle_{\boldsymbol{k}},
\label{eq:C_1}
\end{align}
\begin{align}
    C_2= \left\langle \sum_{i,j=1}^3\hat{d}^2(k_i)\hat{d}^2(k_j)\right\rangle_{\boldsymbol{k}}\geq 1,
    \label{eq:jensen}
\end{align}
\begin{align}
 C_3= \left\langle \hat{\psi}^2_{A_{1g}}(\boldsymbol{k})\sum_{i=1}^3\hat{d}^2(k_i)\right\rangle_{\boldsymbol{k}}
 \label{eq: C_3}
\end{align}
Given the normalization condition \eqref{eq:normalization}, we may prove that $C_2\geq 1$ using Jensen's inequality. 
The values of the constants depend on the particular basis functions. 
In particular, $C_2$ is crucial for the stability of the triplet state, and at the same time a larger $C_2$ reduces the superconducting condensation energy related to the triplet component thereby making it less favorable compared to the singlet component. 
If the singlet component is an $s$-wave with $\hat{\psi}_{A_{1g}}=1$, then $C_1=C_3=1$. 

For given input parameters $\{T_{\mathrm{cs}},T_{\mathrm{ct}}\}$, we would like to know whether the superconducting state is $s$-wave singlet, a triplet $p$-wave BW state, or a singlet-triplet mixed state. For this, we minimize the free energy $F[\psi,\eta]$ with respect to the order parameters $\{\psi,\eta\}$, and compare $F[\psi,0]$ (singlet only), $F[0,\eta]$ (triplet only), and $F[\psi,\eta]$ (mixed) to determine the stable solution.

\subsection{Solutions}

We may write the GL order parameters as $\psi=|\psi|e^{i\varphi_s}$ and $\eta=|\eta|e^{i\varphi_t}$. We must allow for the possibility of a relative phase between the singlet and triplet components $\varphi_s-\varphi_t$. Without loss of generality, we may set $\varphi_t=0$, and minimize for $\varphi_s$. With this, we rewrite the last term of the free energy in Eq. \eqref{eq:free_energy} as $ 2|\psi|^2|\eta|^2\cos(2\varphi_s)$, which in the event of singlet-triplet coexistence, minimizes the free energy for $\varphi_s=\pi/2$. With this, we update the free energy to 
\begin{align}
\frac{1}{2N_0}F[\psi,\eta] & = 
a_s(T)|\psi|^2+ a_t(T)|\eta|^2\notag \\
&+\frac{b(T)}{2}
\left[|\psi|^4+C_2|\eta|^4+2|\psi|^2|\eta|^2
\right],
\end{align}
which we now minimize with respect to $\{|\psi|,|\eta|\}$. 
The constant $C_2$ prevents us from writing the quartic term as a perfect square, which then allows for the possibility of singlet-triplet mixing. 
The solutions for $F[\psi,0]$ and $F[0,\eta]$ are just the textbook solutions \cite{Annett2004}:

\begin{align}
    |\psi_0|^2 = \frac{\ln\left(T_\mathrm{cs}/T\right)}{b(T_\mathrm{cs})}; \quad |\eta_0|^2 = \frac{\ln\left(T_\mathrm{ct}/T\right)}{C_2 b(T_\mathrm{ct})} \label{eq:GL_amplitudes_exclusive}
\end{align}

For the solution $F[\psi,\eta]$ where both components coexist ($|\psi|,|\eta|>0$),  we obtain
\begin{align}
    |\eta|^2 = \frac{\ln\left(T_\mathrm{ct}/T_\mathrm{cs}\right)}{b(T_\mathrm{ct})(C_2-1)};\quad
    |\psi|^2 = \frac{\ln\left[\frac{T}{T_\mathrm{ct}}\left(\frac{T_\mathrm{cs}}{T}\right)^{C_2}\right]}{b(T_\mathrm{ct})(C_2-1)}.
    \label{eq:the_solution}
\end{align}
The analysis of the denominator and the arguments of the logarithms of the
the solution in Eq. \eqref{eq:the_solution} imposes three requirements for coexistence:
\begin{enumerate}
    \item $C_2>1$, which is satisfied for most periodic basis functions. 
    \item $T_\mathrm{ct}>T_\mathrm{cs}$, which means that the triplet instability must be dominant. 
    Therefore, if $T_\mathrm{cs}\geq T_\mathrm{ct}$, there is only a singlet phase, although there is an attractive triplet channel, as would be expected from symmetry arguments.  However, for $T_\mathrm{ct}>T_\mathrm{cs}$, coexistence occurs and remarkably, $|\eta|$ is temperature independent. 
    \item $\frac{T}{T_\mathrm{ct}}\left(\frac{T_\mathrm{cs}}{T}\right)^{C_2}>1$, that is, $|\psi|$ only develops below a secondary transition temperature given by
    \begin{align}
    T_\mathrm{cs}'=\left(\frac{T_\mathrm{cs}}{T_\mathrm{ct}}\right)^{\frac{1}{C_2-1}}T_\mathrm{cs}<T_\mathrm{cs}<T_\mathrm{ct}. 
    \label{eq:axionic_transition}
    \end{align}
\end{enumerate}
Alternatively, Eq. \eqref{eq:axionic_transition} may also be expressed as $T_\mathrm{cs}'=\left(T_\mathrm{cs}/T_\mathrm{ct}\right)^{\frac{C_2}{C_2-1}}T_\mathrm{ct}$.
If the three requirements above are satisfied, then $F[\psi,\eta]$ is the most favorable free energy solution. 

With these analytical results, we may now answer the questions posed at the beginning of Sec. \ref{sec:gl}. If the singlet instability dominates ($T_\mathrm{cs} > T_\mathrm{ct}$), there is only a textbook solution $s$-wave phase. (ii) If $T_\mathrm{cs} = T_\mathrm{ct}$, the singlet state wins, because since $C_2>1$, the triplet state has a lower condensation energy compared to the singlet, rendering it unfavorable. (iii) However, if the triplet instability dominates ($T_\mathrm{ct} > T_\mathrm{cs}$), we find two-phase superconductivity. The first phase occurs in the temperature interval $T_\mathrm{cs}'<T< T_\mathrm{ct}$ for which $F[0,\eta]$ is minimum, and only the triplet state condensates. The second superconducting transition occurs at $T_\mathrm{cs}'$. In the temperature interval $0<T<T_\mathrm{cs}'$, the solution is given by Eq. \eqref{eq:the_solution}, which describes a parity-mixed chiral superconducting $p+is$ state, where inversion and time-reversal are spontaneously broken, hence, an axionic state. This phase is not anticipated from symmetry arguments alone, and required energetics.

\subsection{Phase diagram}

\begin{figure}
\centering
\includegraphics[width = 0.49\textwidth]{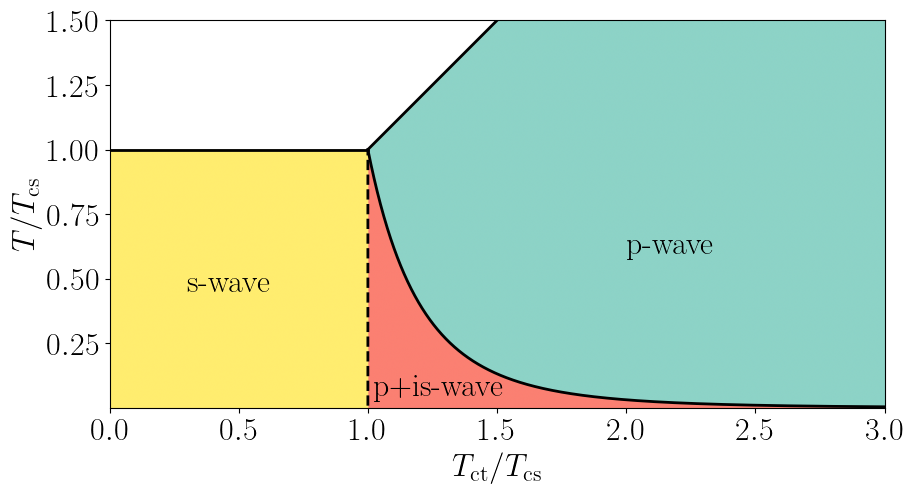}
\caption{
Phase diagram (with $C_2=1.2$) summarizing the three superconducting solutions present in the two-channel model. 
}
\label{fig:polar}
\end{figure}

Fig. \ref{fig:polar} shows a phase diagram with all types of solutions for the input parameters $\{T_{\mathrm{cs}},T_{\mathrm{ct}}\}$ as a function of temperature.
The area of the $p+is$ phase depends on $C_2$, for which we set $C_2=1.2$ for illustration purposes. The line separating the axionic $p+is$ phase from the $p$-wave phase decreases exponentially according to Eq. \eqref{eq:axionic_transition}. The phase diagram shows that the axionic phase is more likely to appear when the triplet instability slightly dominates over the singlet.

In Fig. \ref{fig:specific heat} we show the temperature dependence of the superconducting order parameters for  $T_\mathrm{ct}/T_\mathrm{cs}=1.1$.
The first superconducting transition occurs at $T_\mathrm{ct}$, below which a pure temperature dependent $p$-wave develops (an isotropic BW state). Below $T_\mathrm{cs}'$, the triplet component remains constant, and the additional condensation energy originates from a developing $is$-wave, thereby breaking $\mathcal{P}$ and $\mathcal{T}$, but retaining  $\mathcal{PT}$.

\subsection{Specific heat}

The axionic transition temperature $T_\mathrm{cs}'$ shows up in thermodynamic response functions such as specific heat $c(T)$, which is calculated through the relation \cite{Tinkham2004}
\begin{align}
 c(T) = \frac{2}{NT} \sum_{\boldsymbol{k}}\left(-\frac{\partial f_{\boldsymbol{k}}}{\partial E_{\boldsymbol{k}}}\right)\left( E_{\boldsymbol{k}}^2 -\frac{T}{2}\frac{\partial|\varDelta(\boldsymbol{k})|^2}{\partial T}  \right),
\end{align}
where $f_{\boldsymbol{k}} = (1 + e^{\beta E_{\boldsymbol{k}}})^{-1}$ is the Fermi distribution for the
quasi-particles with energy dispersion $E_{\boldsymbol{k}} = \sqrt{\xi_{\boldsymbol{k}}^2 + |\varDelta(\boldsymbol{k})|^2}$. In the present case, all superconducting states are unitary, such that \cite{Sigrist1991}
\begin{align}
|\varDelta(\boldsymbol{k})|^2=\frac{1}{2}\mathrm{tr}\left[\Delta(\boldsymbol{k})\Delta^\dagger(\boldsymbol{k})\right]=|\psi(\boldsymbol{k})|^2+|\boldsymbol{d}(\boldsymbol{k})|^2.
\end{align}
To calculate the specific heat, we supply
\begin{align}
|\varDelta(\boldsymbol{k})|^2=
\begin{cases}
0,\, \qquad\qquad\qquad\quad \mathrm{if}\quad T\geq T_\mathrm{ct};\\
|\eta_0\hat{\boldsymbol{d}}(\boldsymbol{k})|^2,\,\qquad\,\,\,\quad\mathrm{if}\quad T_\mathrm{cs}'\leq T<T_\mathrm{ct};\\
|\eta\,\hat{\boldsymbol{d}}(\boldsymbol{k})|^2+|\psi|^2, \quad\mathrm{if}\quad 0< T<T_\mathrm{cs}',
\end{cases}
\label{eq:gl_solutions}
\end{align}
where the explicit basis function for the BW triplet state is $ \hat{\boldsymbol{d}}(\boldsymbol{k})=\mathcal{N}\left(\sin(k_xa),\sin(k_ya),\sin(k_za)\right)$, and $\mathcal{N}$ is the normalization constant fixed by Eq. \eqref{eq:normalization}. 

\begin{figure}
    \centering
\includegraphics[width=0.49\textwidth]{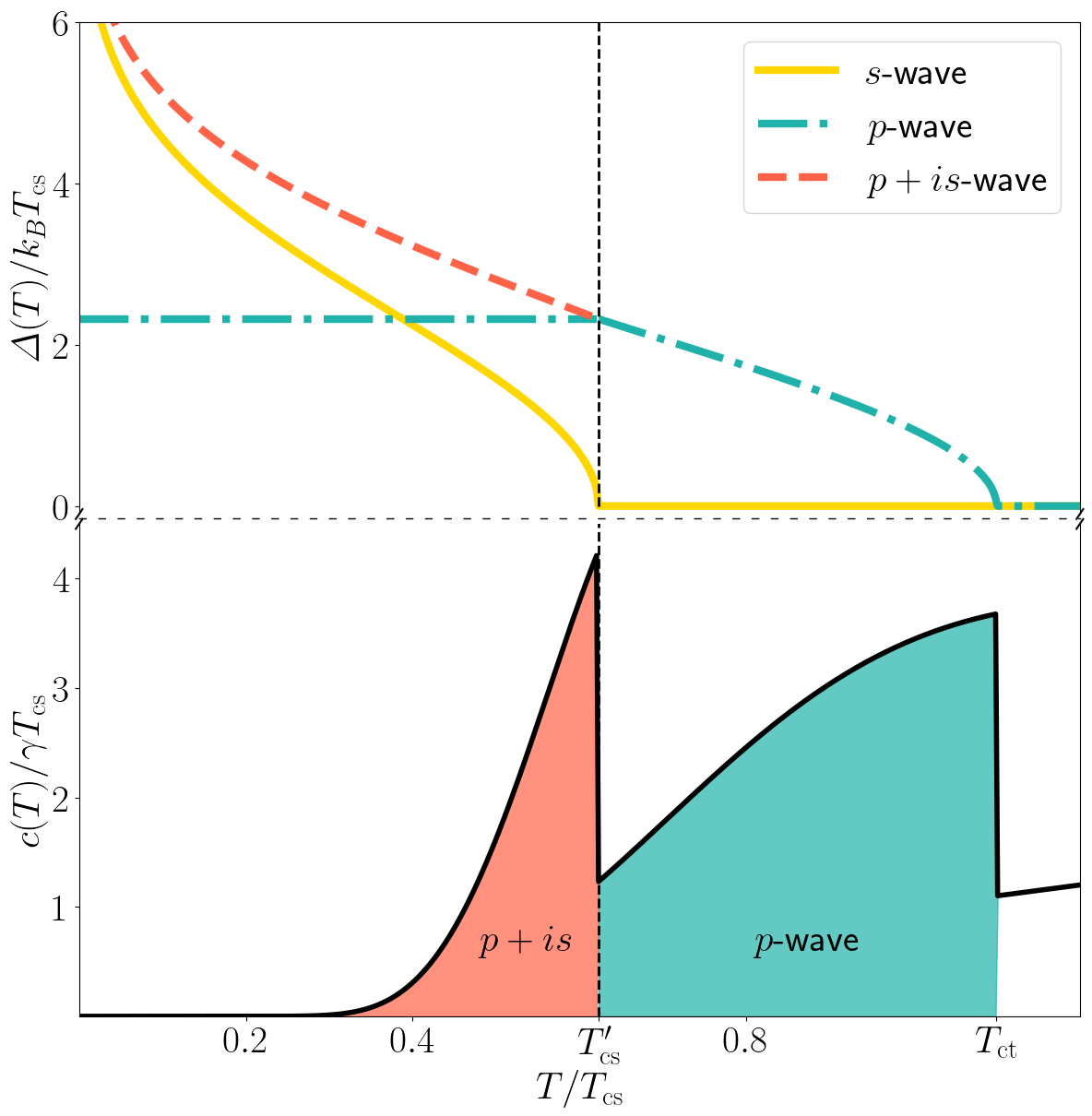}
\caption{
Temperature dependence of the Ginzburg-Landau order parameters and specific heat for $T_\mathrm{ct}/T_\mathrm{cs} = 1.1$ and $C_2=1.2$.
}
\label{fig:specific heat}
\end{figure}

We show the plot of the specific heat in Fig.~\ref{fig:specific heat}, using the same parameters as in Fig.~\ref{fig:polar}. The normal state is described by the specific heat of a Fermi gas, $c_0(T) = \gamma T$, with $\gamma = 4\pi^2 N_0 / 3$. The two superconducting transitions manifest as a double transition in the specific heat, similar to what occurs in chiral superconductors~\cite{Roising2022}.  
The crucial difference here, compared to chiral superconductors, is that inversion symmetry is broken together with time-reversal symmetry.  
The magnitude of the specific heat discontinuity at $T_\mathrm{ct}$ is given by $\Delta c(T_\mathrm{ct}) / c_0 = 1.43 / C_2$.  
Since we used the Ginzburg-Landau order parameters as input to Eq.~\eqref{eq:gl_solutions}, the slopes in $c(T)$ should be regarded as qualitative.

Some examples where double transitions in specific heat have been reported include UPt$_3$ \cite{Fisher1989}, PrOs$_4$Sb$_{12}$ \cite{Braithwaite2008}, Sr$_2$RuO$4$ \cite{Roising2022}, U$_{1-x}$Th$_x$Be$_{13}$ \cite{Stewart2019}, and UTe$_2$ \cite{Hayes2021,Rosa2022}. 
While many possible origins exist for double specific heat transitions, and we do not focus on a specific material in this work, the observation of such transitions motivates future investigations.


\section{The Hubbard-Holstein model \label{sec:mechanisms}}

As is common in BCS-type theories and group-theoretical methods \cite{Sigrist1991}, the existence of attractive channels is typically assumed. This approach is popular because it bypasses the complexities of pairing mechanisms and suffices for most practical purposes. We adopted this same approach in the Hamiltonian of Eq. \eqref{eq:bcs}, which led to the axionic $p+is$ state.

However, to explain superconductivity and identify candidate materials with
two attractive channels, one must start with a model Hamiltonian that includes specific interactions to properly populate — and thus justify — the effective pairing channels in Eq. \eqref{eq:channels}.
In this section, we critically examine whether a microscopic model containing both ferromagnetic spin fluctuations and phonons could support two attractive pairing channels.

One immediate limitation is that the two-channel axionic superconductor must exhibit a dominant triplet instability. To the authors' knowledge, among the less controversial triplet superconductor candidates are uranium-based superconductors \cite{Mineev2014,Mineev2017}, which are typically ferromagnetic or near a ferromagnetic instability. As motivation, while admittedly not (yet) committing to specific materials, we consider a model close to a ferromagnetic instability. Any such system also features electron-phonon coupling, which may or may not provide a subdominant singlet channel.

\subsection{The model}

We now study the Hubbard-Holstein model \cite{vonderLinden1995}, and examine how its effective theory populates the pairing interactions in Eq. \eqref{eq:channels}. The model is given by
\begin{align}
\mathcal{H}=& \sum_{i,j,s}t_{ij}c_{is}^\dag c_{js}-\mu\sum_{i,s}n_{is} +U\sum_i n_{i\uparrow}n_{i\downarrow} \notag \\
&+\omega_0\sum_i b_i^ \dag b_i+g\sum_{i,s}n_{is}\left(b_i^ \dag+b_i\right),
\label{eq:hh}
\end{align}
where $t_{ij}$ are the inter-site hoppings, $\mu$ is the chemical potential, $n_{is}=c^\dag_{is}c_{is}$ is the electron number operator for a spin $s=\{\uparrow,\downarrow\}$ at site $i$, $U$ is the on-site Hubbard Coulomb repulsion, $\omega_0$ is a localized phonon mode with $b_i^\dag$ and $b_i$ being the phonon creation and annihilation operators, and $g$ is an electron-phonon Holstein coupling. 

The Hubbard-Holstein model exhibits a diverse range of Fermi surface instabilities, including superconductivity, spin and charge instabilities \cite{vonderLinden1995,Costa2020}. Here, we focus specifically on the $s$-wave superconducting channel, which may arise from the electron-phonon coupling $g$, and the $p$-wave superconducting channel, driven by ferromagnetic spin fluctuations induced by the repulsive interaction $U$.

\subsection{Derivation of the effective model}

We use the path integral approach to formulate the partition function
\begin{align}
\mathcal{Z}=\int\mathcal{D}\left(\bar{\psi},\psi\right)e^{-\mathcal{S}},
\end{align}
where the action is expressed in terms of the Hamiltonian as
\begin{align}
 \mathcal{S}=\int_0^\beta \mathrm{d}\tau\left[\sum_{i,s}\bar{\psi}_{is}(\tau)\frac{\partial}{\partial \tau}\psi_{is}(\tau)+\mathcal{H} \right ] .
\end{align}
The operators $\{c_{is}^\dag,c_{is}\}$ are expressed in terms of their Grassmann counterparts $\{\bar{\psi}_{is},\psi_{is}\}$, $\tau$ is imaginary time, and $\beta=1/T$ is the inverse temperature.

The Hubbard-Holstein model is translationally invariant. For this reason, we Fourier transform the Grassmann variables to momentum-frequency space using
\begin{align}
\psi_{js}(\tau) & =\frac{1}{\sqrt{N\beta}}\sum_{\boldsymbol{k},\omega_n}\psi_{\boldsymbol{k}s}(i\omega_n)e^{-i\left(\omega_n\tau-\boldsymbol{k}\cdot\boldsymbol{R}_j\right)}; \notag \\
\bar{\psi}_{js}(\tau) & =\frac{1}{\sqrt{N\beta}}\sum_{\boldsymbol{k},\omega_n}\bar{\psi}_{\boldsymbol{k}s}(i\omega_n)e^{i\left(\omega_n\tau-\boldsymbol{k}\cdot\boldsymbol{R}_j\right)},
\end{align}
where $N$ is the number of lattice sites, and $\omega_n=(2n+1)\pi/\beta$ are the fermionic Matsubara frequencies.

Next, we consider the ferromagnetic spin fluctuations arising from the Hubbard $U$ term. These fluctuations, together with the phonons, constitute the bosonic degrees of freedom. By integrating out these bosonic degrees of freedom, we obtain an effective fermionic Hamiltonian that maps to Eq. \eqref{eq:channels}.
Below, we summarize the main rationale behind the procedure.

\subsubsection{Electron-phonon interaction}

We follow the standard textbook treatment of the electron-phonon interaction, making simplifying assumptions that reduce it to the BCS pairing form \cite{Annett2004,Altland2023}.
Within the path-integral formalism, this involves shifting the bosonic variables, followed by integrating out the bosonic degrees of freedom. By considering a single phonon branch characterized by a single phonon mode with frequency $\omega_0$, the effective density-density interaction reads (converting back from Grassmann variables to operators):
\begin{align}
-V\sum_{\boldsymbol{k},\boldsymbol{k}'} 
c_{\boldsymbol{k}\uparrow}^\dag c_{-\boldsymbol{k}\downarrow}^\dag c_{-\boldsymbol{k}'\downarrow}c_{\boldsymbol{k}'\uparrow},\quad V=\frac{g^ 2}{\omega_0},
\label{eq:dd}
\end{align} 
which is attractive in the $s$-wave singlet channel. However, in the presence of a repulsive $U$, a transformation back to real space shows that the role of the electron-phonon interaction is to rescale the on-site repulsion \cite{Capone2010}, such that the Hubbard-Holstein model with integrated out phonons reads
\begin{align}
\tilde{\mathcal{H}}= \sum_{i,j,s}t_{ij}c_{is}^\dag c_{js}-\mu\sum_{i,s}n_{is} +\tilde{U}\sum_i n_{i\uparrow}n_{i\downarrow},
\end{align}
where $\tilde{U}=U-V$. Depending on the sign of $\tilde{U}$, this is a repulsive or attractive Hubbard model.

\subsubsection{Ferromagnetic spin fluctuations}

We next assume $\tilde{U}>0$, and that the conditions for proximity to itinerant ferromagnetism are satisfied \cite{Moriya1985,Coleman2015}.
To look at the ferromagnetic channel of the Hubbard part, we introduce the magnetization $\boldsymbol{M}_{\boldsymbol{q}}(i\nu_m)$ as the Hubbard-Stratonovich field, which fluctuates about its mean field value $\boldsymbol{M}$:
\begin{align}
 \boldsymbol{M}_{\boldsymbol{q}}(i\nu_m)=\boldsymbol{M}\delta_{\boldsymbol{q},0}\delta_{\nu_m,0}+\delta \boldsymbol{M}_{\boldsymbol{q}}(i\nu_m),
\end{align}
where $\nu_m=2m\pi/\beta$ are the bosonic Matsubara frequencies. We will ultimately be interested in the fluctuating paramagnetic region, where $\boldsymbol{M}=0$. Considering Gaussian fluctuations and integrating them leads to an effective spin-spin interaction of the form \cite{Nakajima1973,Coleman2015,Sigrist2005,Scalapino2012,Sachdev2023}
\begin{align}
\frac{1}{2}\sum_{\boldsymbol{k},\boldsymbol{k}'}\sum_{\{s_i\}}J_{\boldsymbol{k}-\boldsymbol{k}'}\,\boldsymbol{\sigma}_{s_1s_1'}\cdot\boldsymbol{\sigma}_{s_2s_2'}\, c_{\boldsymbol{k},s_1}^\dag c_{-\boldsymbol{k},s_2}^\dag c_{-\boldsymbol{k}',s_2'}c_{\boldsymbol{k}',s_1'},
\label{eq:ss}
\end{align} 
where the effective interaction is given by
\begin{align}
J_{\boldsymbol{q}}=-\frac{I}{1-I\chi_0(\boldsymbol{q})},\quad I=\frac{\tilde{U}}{3},
\end{align}
where $\chi_0(\boldsymbol{q})$ is the bare static paramagnetic susceptibility. Close to a ferromagnetic instability the interaction $J_{\boldsymbol{q}}$ enhances favoring a superconducting triplet channel. 

Assuming a parabolic dispersion for the normal state,
the paramagnetic susceptibility may be expanded for small momenta for which 
\cite{Sigrist2005}
\begin{align}
 \chi_0(\boldsymbol{q})\approx N_0\left(1-\frac{q^ 2}{12k_\mathrm{F}^ 2}\right),
\end{align}
where $N_0$ is the density of states at the Fermi level, and $k_\mathrm{F}$ is the Fermi momentum. Then, we may also expand the effective interaction as
\begin{align}
 J_{\boldsymbol{q}}\approx J_0\left[1-\frac{N_0J_0}{12}\left(\frac{q}{k_\mathrm{F}}\right)^ 2\right],
\end{align}
where $J_0=J_{\boldsymbol{q}=0}=I/(1-IN_0)$.

\subsubsection{The effective model}

Given the effective spin-spin interaction obtained in Eq. \eqref{eq:ss}, we write the effective Hamiltonian as \cite{Azambuja2025}
\begin{align}
& \mathcal{H}_\mathrm{eff}=\sum_{\boldsymbol{k},s}\xi_{\boldsymbol{k}}c_{\boldsymbol{k}s}^\dag c_{\boldsymbol{k}s} \notag \\
&+
\frac{1}{2}\sum_{\boldsymbol{k},\boldsymbol{k}'}\sum_{\{s_i\}}V_{s_1s_2,s_1's_2'}(\boldsymbol{k}-\boldsymbol{k}')\, 
c_{\boldsymbol{k},s_1}^\dag c_{-\boldsymbol{k},s_2}^\dag c_{-\boldsymbol{k}',s_2'}c_{\boldsymbol{k}',s_1'},
\end{align} 
with the effective interaction given by 
\begin{align}
V_{s_1s_2,s_1's_2'}(\boldsymbol{k}-\boldsymbol{k}') =J_{\boldsymbol{k}-\boldsymbol{k}'}\,\boldsymbol{\sigma}_{s_1s_1'}\cdot\boldsymbol{\sigma}_{s_2s_2'}.
\label{eq:eff_v}
\end{align} 
The interaction in Eq. \eqref{eq:eff_v} is not yet fully symmetrized into singlet and triplet channels as in Eq. \eqref{eq:channels}. 
The spin-spin interaction populates both singlet and triplet channels, which may be attractive or repulsive.

\subsection{Pairing channels}

To finalize the mapping of Eq. \eqref{eq:eff_v} to the form of Eq. \eqref{eq:channels}, we symmetrize the effective interaction in Eq. \eqref{eq:eff_v} using the procedures described in Refs. \cite{Sigrist2005,Samokhin2008,Sigrist2009,Coleman2015}. 
There is some algebra involved in the procedure. 
However, the singlet and triplet contributions of the spin-spin interaction $J=J_s+J_t$ may be recognized via
\begin{align}
J_s(\boldsymbol{k},\boldsymbol{k}')& = \frac{3}{4}\left(J_{\boldsymbol{k}-\boldsymbol{k}'}+J_{\boldsymbol{k}+\boldsymbol{k}'}\right)  \notag \\
&
= \frac{3}{2}J_0\left[1-\frac{N_0J_0}{12k_\mathrm{F}^2}\left(|\boldsymbol{k}|^2+|\boldsymbol{k}'|^2\right)\right],
\end{align}
\begin{align}
J_t(\boldsymbol{k},\boldsymbol{k}')& = - \frac{1}{4}\left(J_{\boldsymbol{k}-\boldsymbol{k}'}-J_{\boldsymbol{k}+\boldsymbol{k}'}\right)
=-\frac{N_0 J_0^2}{24k_\mathrm{F}^2}\boldsymbol{k}\cdot\boldsymbol{k}'.
\end{align}
The singlet spin-spin interaction contains a repulsive part $\sim 1$ in the $s$-wave channel and an attractive part  $\sim |\boldsymbol{k}|^2=k_x^2+k_y^2+k_z^2$ in an extended $s$-wave channel. The spin-spin interaction also populates an attractive triplet channel $\sim \boldsymbol{k}=(k_x,k_y,k_z)$,
which corresponds to a $p$-wave channel. 
The mapping of the Hubbard-Holstein model, as defined in Eq.~\eqref{eq:hh}, with a bias towards ferromagnetic spin fluctuations, to the effective model described in Eq.~\eqref{eq:bcs}, is summarized in Table~\ref{tab:oh}. 

\begin{table}
\begin{tabular}{lllll}
\hline
Irrep  & Basis function      & Coupling               & Name              & Type     \\\hline  \\[-1ex]
$A_{1g}$ & $1$                 & $\frac{3J_0}{2}$                  & $s$-wave          & Repulsive  \\
$A_{1g}$ & $k_x^2+k_y^2+k_z^2$ & $-\frac{N_0J_0^2}{12}$ & $s'$-wave & Attractive \\
$T_{1u}$ & $(k_x,k_y,k_z)$     & $-\frac{N_0J_0^2}{24}$   & $p$-wave          & Attractive \\[1ex] \hline
\end{tabular}
\caption{Pairing channels of the effective model.}
\label{tab:oh}
\end{table}

For illustrative purposes, we again consider a cubic system with point group symmetry \( O_h \) and identify the active channels of the effective model. 
Of the ten possible irreps of $O_h$, the model only contributes
to the \( A_{1g} \) and \( T_{1u} \) channels. 
Within the \( A_{1g} \) channel, ferromagnetic spin-fluctuations suppress the conventional \( s \)-wave pairing while favoring an extended \( s' \)-wave component. Additionally, the spin fluctuations populate the parity-odd \( T_{1u} \) channel, leading to an attractive \( p \)-wave.
We emphasize that more channels would be populated in the case of a non-quadratic dispersion; however, the main point is that the $s$-wave channel remains repulsive. 

Thus, the Hubbard-Holstein model allows for a possible dominant triplet instability in the $T_{1u}$ channel. However, the $s$-wave channel is repulsive, ruling out a $p+is$ state within the present assumptions. Nevertheless, a subdominant attractive extended $s'$-wave $A_{1g}$ channel exists. The ratio of the triplet $T_{1u}$ and extended singlet $A_{1g}$ couplings is $v_{t,T_{1u}}/v_{s',A_{1g}} = 2$, opening the possibility for an axionic $p+is'$ state. 


\section{Conclusion \label{sec:discussion}}

In Sec. \ref{sec:gl}, we obtained an axionic superconducting state from two mechanisms, which requires that the triplet channel dominates over the singlet.
However, in Sec. \ref{sec:mechanisms}, we demonstrated that the assumption of two attractive channels is not trivially justified. Therefore, it may be worthwhile to explore alternative theoretical frameworks, such as extended Hubbard models and Kondo channel Hamiltonians.

We found that if the singlet $s$-wave instability dominates, there is no need to consider subdominant triplet instabilities. As expected from symmetry arguments, in the singlet-dominated case, the singlet and triplet order parameters are mutually exclusive; however, the converse does not hold. If there are two attractive channels (an $s$-wave singlet and a triplet) and the triplet channel dominates, both inversion and time-reversal symmetries are spontaneously broken by the superconducting state, resulting in an axionic $p+is$ phase. No spin-orbit coupling is required. 
This state is accompanied by a range of experimental signatures characteristic of symmetry breaking superconductors, among which we highlight the double transition in the specific heat.


\begin{acknowledgments}
We thank Sergio Magalhães, Eleonir Calegari, Gerardo Pino and Paulo Pureur for valuable discussions. The authors acknowledge the support from CNPq, FAPERGS and CAPES.
\end{acknowledgments}


\twocolumngrid

\bibliography{references}

\end{document}